\documentclass[a4paper,10pt]{iopart}
\usepackage{iopams}  
\usepackage[colorlinks,citecolor=blue,linkcolor=red,anchorcolor=blue,filecolor=blue,urlcolor=blue]{hyperref}
\usepackage{color}
\usepackage{graphicx}
\usepackage{cite}

\begin{document}
%
\title{Nuclear level densities away from line of $\beta$-stability}
\bigskip
\author{T. Ghosh$^{1,2}$, B. Maheshwari$^{3,\star}$, Sangeeta$^4$, G. Saxena$^5$, B. K. Agrawal$^{1,2,\dagger}$}

\address{$^1$Saha Institute of Nuclear Physics, Kolkata-700064, India}

\address{$^2$Homi Bhabha National Institute, Anushakti Nagar, Mumbai-400094, India}

\address{$^3$Department of Physics, Indian Institute of Technology Ropar, Rupnagar-140001, India}

\address{$^4$Department of Applied Sciences, Chandigarh Engineering College, Landran-140307, India}

\address{$^5$Department of Physics (H $\&$ S), Govt. Women Engineering College, Ajmer-305002, India}


\ead{bhoomika.physics@gmail.com$^\star$;bijay.agrawal@saha.ac.in$^\dagger$} 
\begin{abstract}
\noindent
The variation of total nuclear level densities (NLDs) and level density parameters with proton number $(Z)$ are studied around the $\beta$-stable isotope, $Z_{0}$, for a given mass number. We perform our analysis for a mass range $A=40$ to 180 using the NLDs from popularly used databases obtained with the single-particle energies from two different microsopic mass-models. These NLDs which include microscopic structural effects such as collective enhancement, pairing and shell corrections, do not exhibit inverted parabolic trend with a strong peak at $Z_{0}$ as predicted earlier.  
We also compute the NLDs using the single-particle energies from macroscopic-microscopic mass-model. Once the collective and pairing effects are ignored, the inverted parabolic trends of NLDs and the corresponding level density parameters become somewhat visible. Nevertheless, the factor that governs the $(Z-Z_{0})$ dependence of the level density parameter, leading to the inverted parabolic trend, is found to be smaller by an order of magnitude. We further find that the $(Z-Z_{0})$ dependence of NLDs is quite sensitive to the shell effects.
\end{abstract}
\section{Introduction}
The total nuclear level density (NLD) is a fundamental quantity required to study various aspects of nuclear physics \cite{Ericson1960,Huizenga1972,Karampagia2020,firestone2021}. The NLDs are one of the key ingredients which enter into the calculations of reaction cross-sections \cite{Hauser1952,Arnould2007}. The evaluation of astrophysical reaction rates at a fixed temperature requires Maxwellian average of cross sections over a wide range of energy \cite{claus1989}. Experimental data on NLDs are mostly available for the nuclei near stability line \cite{oslo}. However, the NLDs for the nuclei away from the line of $\beta$-stability are crucial inputs for understanding the astrophysical processes and one has to rely mostly on theoretical estimates. The different theoretical estimates near drip-line tend to differ significantly leading to large uncertainties \cite{Liddick2016}. \par

Many theoretical and experimental efforts are made to estimate NLDs of nuclei away from line of stability \cite{Brondi2010,Voinov2012,Bezbakh2015,Liddick2016,Roy2020}. Usually, one extrapolates the level densities beyond the stability line on the basis of known values. This is achieved by modifying the values of level density parameter $a$ in the Bethe's formula of level density based on the Fermi gas model \cite{bethe1936, bethe1937}. For a given mass, the dependencies of the level density parameter on neutron-proton difference $(N-Z)$, or on $(Z-Z_{0})$ with $Z_{0}$ being the proton number for the most stable nucleus, are explored in Refs. \cite{AlQuraishi2001,AlQuraishi2003,Grimes1990,grimes2002,charity2003}. Such modifications in level density parameter lead to marginal improvements in the fit to the available data of NLDs for masses $20\le A \le 110$; the $\chi^2$ function improves only by the order of $\sim10\%$. The data are mostly available for the nuclei around  $|Z-Z_{0}|\le 1$, beyond which the data are exiguous. 
The fit to these data when extrapolated to $|Z-Z_{0}| > 1$ suggests strong suppression of level density or level density parameter. For instance, the level density parameter reduces by a factor of 2 for $(Z-Z_{0}) = 4$ which amounts to reduction in level density by $\sim 10^{4}$ for $A = 200$ at excitation energy $E^{*} = 10$ MeV \cite{AlQuraishi2003}.

Recently, there are some measurements on the level densities for the nuclei with $|Z-Z_{0}| > 1$. The measured NLDs for $^{70}$Ni $(Z-Z_{0}=3)$ \cite{Liddick2016} and $^{115}$Te  $(Z-Z_{0}=2)$ \cite{Roy2020} manifest that the strong reduction in NLDs as predicted by AlQuraishi $et$ $al.$ \cite{AlQuraishi2003} are overestimated. This suggests that the extrapolation beyond $|Z-Z_0|=1$ needs to be re-examined. The $(Z-Z_0)$ dependence of the level densities may be quite sensitive to the single-particle energies which may change drastically as one moves away from the $\beta$-stability line. Consequently, the other microscopic effects such as collective enhancement due to rotational and vibrational states  \cite{Gilbert1965,Dilg1973,dossing1974,Huizenga1974,Ignatyuk1979,Ignatyuk1985,pandit2018,dossing2019,mohanto2019}, pairing and shell corrections \cite{Ignatyuk1968,decowski1968,Rubchenya1970,Ramamurthy1972,Schmidt1982,agrwal1999,Schmidt2012} will also be crucial in determining the behavior of level density around $Z_0$ \cite{Gilbert1965, Bezbakh2015, Guttormsen2021}. One of the ways to include various microscopic effects in the level density parameter is through the semi-classical approach. The trace formula in this approach takes care of the oscillatory part of the level density and it has successfully been applied to study the influence of damping and melting of shell effects in the level density parameter \cite{Kaur2015,Dwivedi2019}. The level densities are also studied extensively using various microscopic approaches based on 
Hartree-Fock \cite{Demetriou2001} and Hartree-Fock-Bogoliubov approximations \cite{Goriely2008}.  

In the present work, we investigate the $(Z-Z_{0})$ dependence of level density and level density parameter for a wide range of masses $40 \le A \le 180$. We analyze the level densities obtained from databases corresponding to two different microscopic mass-models, namely, Hartree-Fock  with  pairing  treated  within  the  Bardeen-Cooper-Schrieffer  approximation  (HF) \cite{Demetriou2001}  and Hartree-Fock-Bogoliubov  (HFB) \cite{Goriely2008} using Skyrme forces. These NLDs at low excitation energies are computed using combinatorial method \cite{Egorov1989,Egorov1989npa} which are further normalized with experimental data at low energy and neutron separation energy. The statistical method is used for the computation of NLDs at higher excitation energies. We also compute the level densities using statistical method for the single-particle energies taken from macroscopic-microscopic mass-model, Finite Range Droplet Model (FRDM) \cite{moller1995}. The influence of several microscopic effects on the $(Z-Z_{0})$ dependence of the level density is analyzed.

\section{Statistical model for nuclear level densities}

The nuclear level densities for a given set of discrete single-particle energies can be computed reliably using the statistical model particularly at high excitation energies. In statistical model, one starts with defining the grand canonical partition function for a given nucleus at a fixed temperature. The total nuclear level densities are obtained by the saddle point approximation to the inverse Laplace transformation of the grand canonical partition function. 
It may be pointed out that the saddle point approximation breaks down at a very low temperature which can be taken care by employing a special mathematical technique to avoid
divergence in the limit $T \rightarrow 0$ as discussed in refs.\cite{Tanabe1981,Sugawara1981}. Alternatively, the level density at low excitation energies can be calculated using combinatorial method. In the present work, we consider the level densities at the excitation energies corresponding to the temperature T $\sim1-2$ MeV for which the saddle point approximation is reliable.
For a system of independent nucleons, the grand canonical partition function can be written as,
\begin{figure*}[!htb]
\begin{center}
\includegraphics[width=5.0in, height=4.0in]{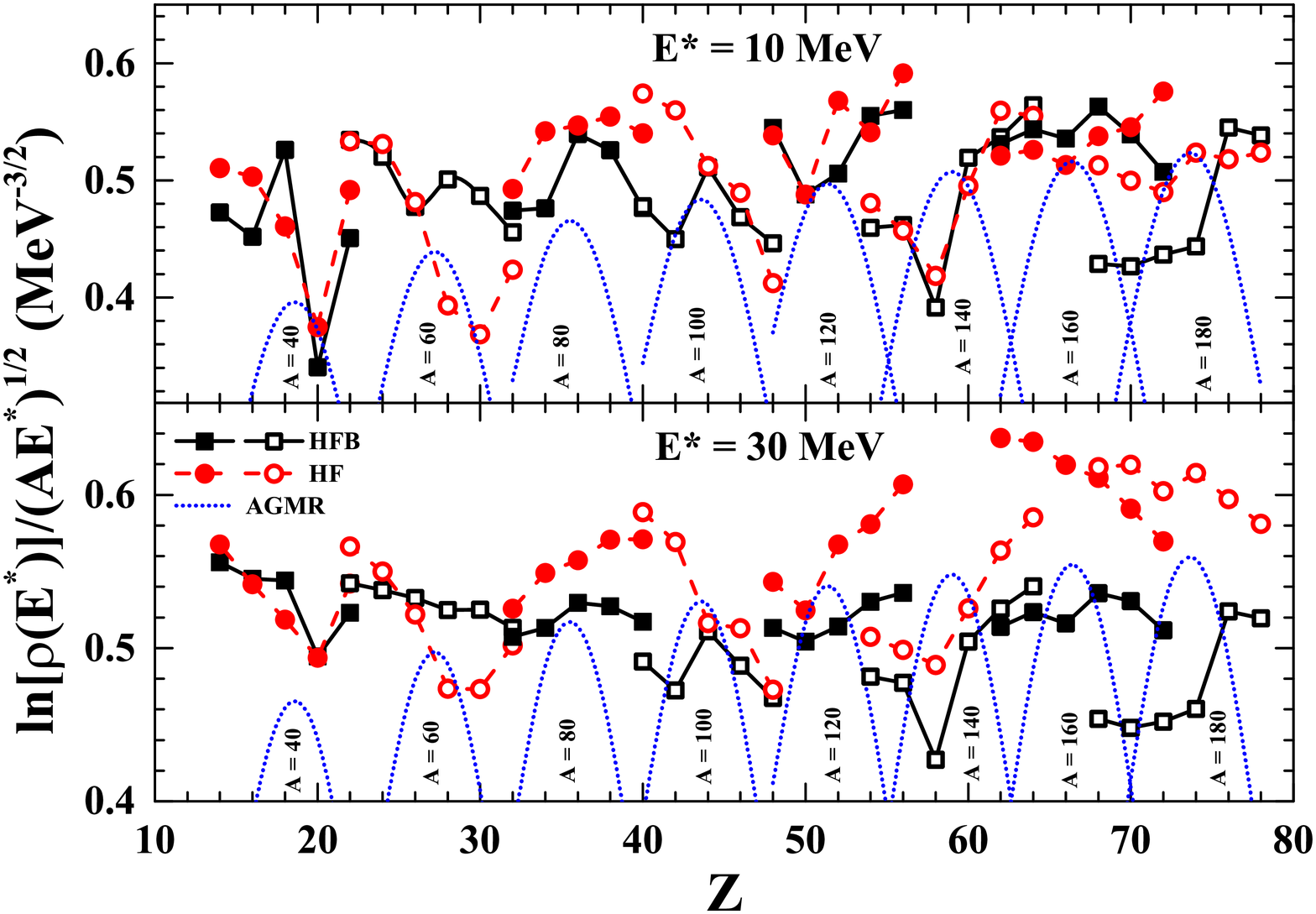}
\caption{The variation of scaled NLDs as a function of proton number $(Z)$ around the $\beta$-stable isotope $Z_0$ for fixed mass. The results are shown for $A = 40-180$ at the excitation energies $E^*=10$ and 30 MeV. The NLDs are taken from HF and HFB databases which include the effects of collective enhancement, pairing and shell structure. For the comparison, NLDs obtained with AGMR form of level density parameter defined in Eq. (\ref{ldzo}), are also shown.}\label{fig:hf}
\end{center}
\end{figure*} 

\begin{figure*}[!htb]
\begin{center}
\includegraphics[width=5.0in, height=4.0in]{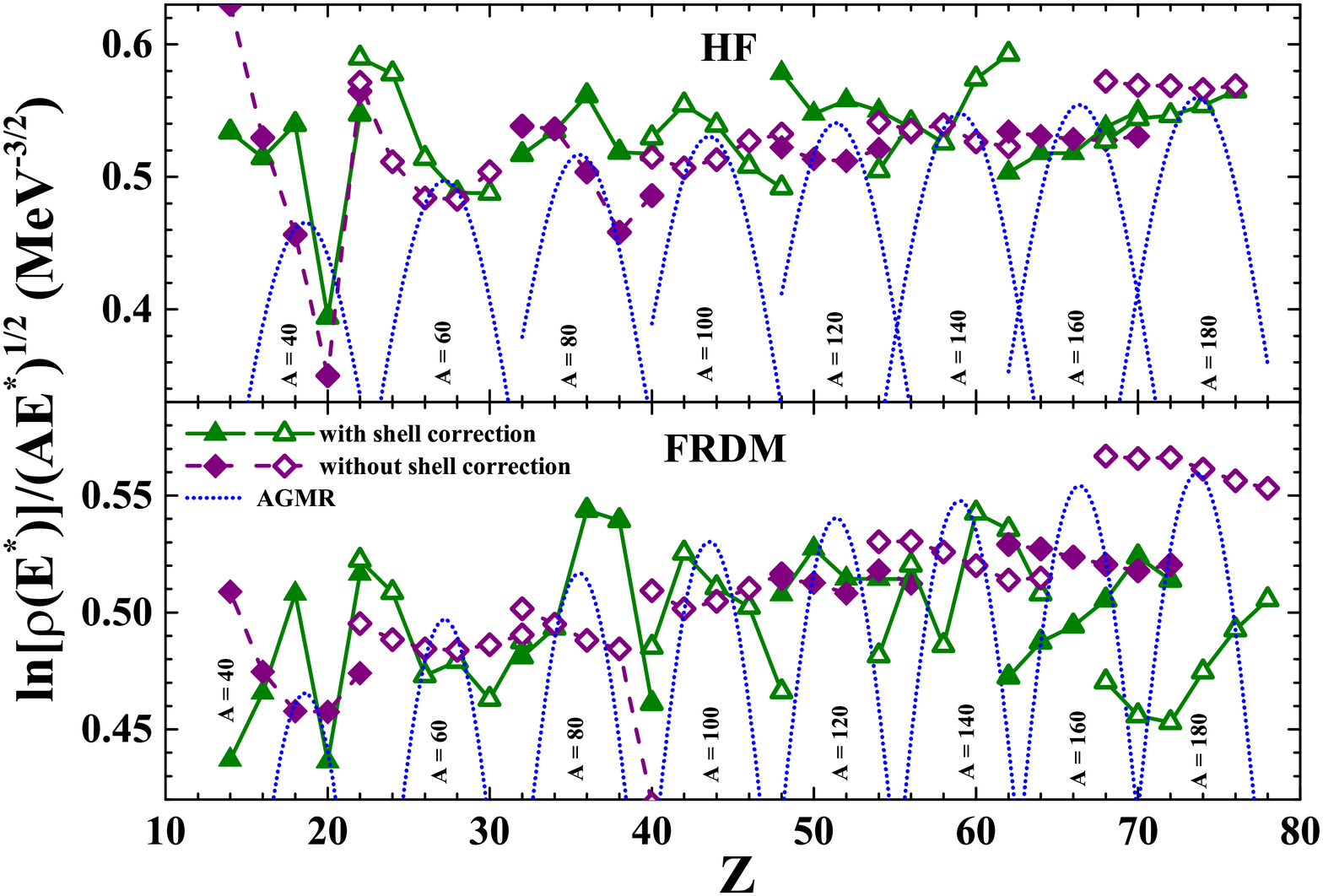}
\vspace{-0.2cm}
\caption{The scaled NLDs at $E^*=30$ MeV with and without shell corrections at and around $Z_0$ for fixed $A$ using HF and FRDM single-particle energies. These NLDs are obtained by ignoring the collective enhancement and pairing effects.}\label{fig:frdm}
\end{center}
\end{figure*} 
\begin{equation}
    \mathcal{Z}(T,\mu_n,\mu_p)=Tr ~{e^{-\frac{1}{T}(E-\mu_nN- \mu_p Z)},}
\end{equation}
where $N$ and $Z$ are neutron and proton numbers. The $\mu_n$ and $\mu_p$ are chemical potentials for neutrons and protons, respectively. Here, T, $\mu_n$ and $\mu_p$ are related to the energy, neutron and proton numbers as,
\begin{eqnarray}
    E={T^2}\frac{\partial}{\partial T} ln \mathcal{Z} +{\mu_n N}+{\mu_p Z} ,\\
    N= T\frac{\partial}{\partial \mu_n} ln \mathcal{Z} ,\\
    Z= T\frac{\partial}{\partial \mu_p} ln \mathcal{Z}.
\end{eqnarray}
Once the $\mu_n$ and $\mu_p$ are adjusted to reproduce the required nucleon numbers at a given temperature $(T)$, the entropy $S$ can be obtained from the following equation as,
\begin{eqnarray}
-Tln \mathcal{Z}=E-TS-{\mu_n N}-{\mu_p Z}.
\end{eqnarray}
The level density in the saddle point approximation can now be written as,
\begin{eqnarray}
\rho(E,N,Z)=\frac{1}{(2 \pi)^{3/2}} \frac{e^S}{D^{1/2}},
\label{ld}
\end{eqnarray}
where $D$ is the determinant of $3\times3$ matrix whose matrix elements are defined as,
\begin{eqnarray}
d_{ij}=\frac{\partial^2}{\partial x_i \partial x_j}ln \mathcal{Z},
\end{eqnarray}
with the indices $i,j=1,2,3$, and $x_1=\frac{1}{T}$, $x_2=\frac{\mu_n}{T}$, $x_3=\frac{\mu_p}{T}$.
We can calculate level density by using Eq. (\ref{ld}) at excitation energy $E^{*}=E(T)-E(0)$. 

\section{$(Z-Z_{0})$ dependence of nuclear level densities}

Level density for a given nucleus is sensitive to the choice of single-particle energies. It becomes increasingly important to employ the appropriate set of single-particle energies in order to study the $(N-Z)$ or $(Z-Z_{0})$ dependence of NLDs at a fixed mass number $A$. There are several microscopic mass-models based on HF \cite{tondeur2000, goriely2001hf} and HFB \cite{samyn2002} applied to Skyrme forces. The parameters of these mass-models are fitted to the measured binding energies of large number ($\sim$2000) of nuclei which also span the region away from the line of $\beta-$stability. Both the HF and HFB mass-models yield comparable fits to the binding energies. Another popularly used mass-model is the FRDM \cite{moller1995} based on so-called macroscopic-microscopic approach. The microscopic part of FRDM is the shell and pairing corrections obtained for single-particle energies from folded-Yukawa potential. \par
 
We first consider the compilations of NLDs obtained for HF and HFB mass-models \cite{Capote2009}. These NLDs at low excitation energies are computed using combinatorial method. They are further normalized with experimental data at low energy and neutron separation energy. The statistical method is used for the computation of NLDs at higher excitation energies. The enhancement of level densities due to collective rotations and vibrations are also incorporated in these NLD databases. We study these level densities for a wide range of masses $40 \le A \le 180$ at $E^*=10$ and 30 MeV. To make these NLDs more or less independent of mass and excitation energy, we use an appropriate factor on the basis of Bethe's formula which is given as,

\begin{figure}[!htb]
\begin{center}
\includegraphics[width=5.0in, height=4.0in]{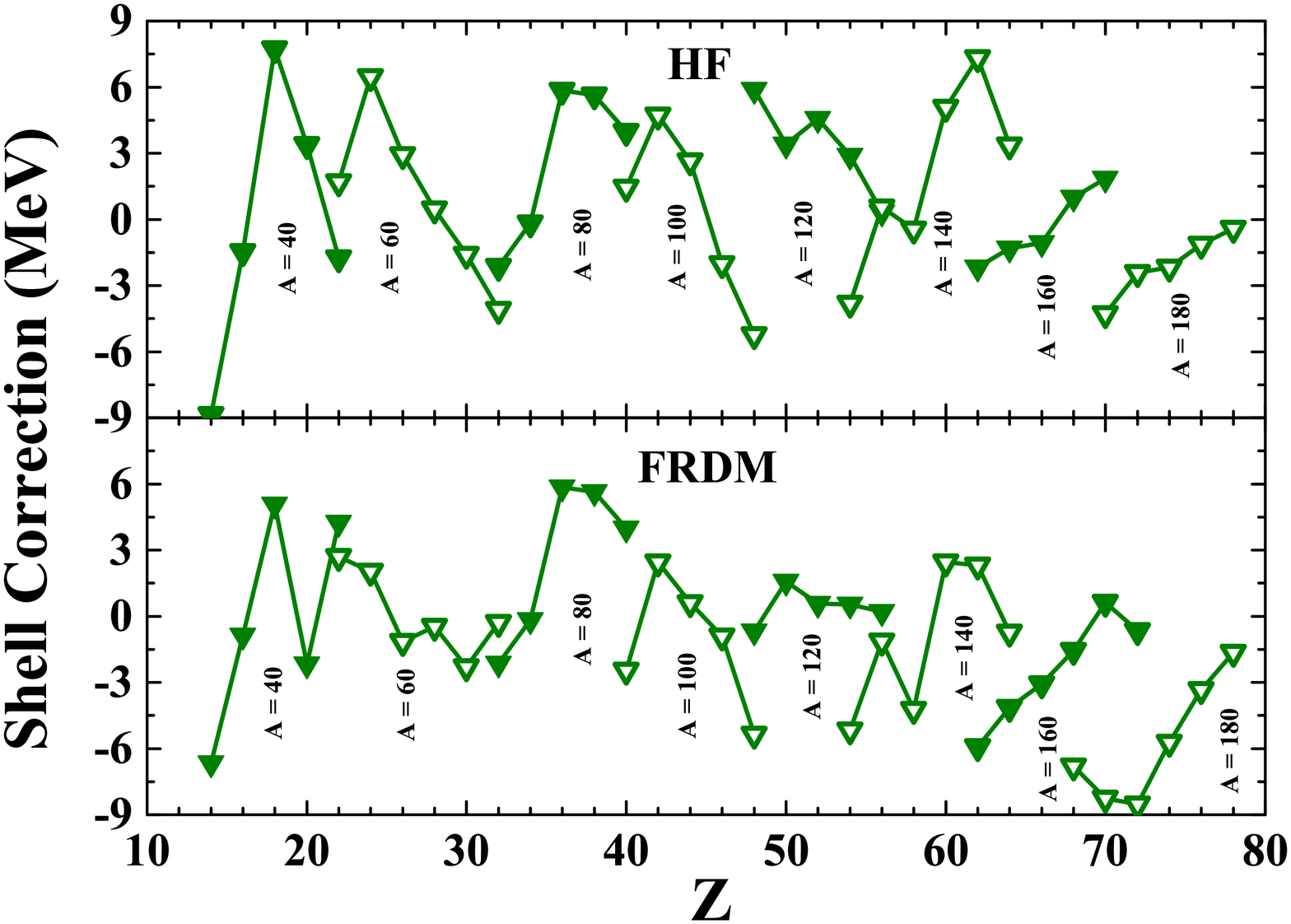}
\vspace{-0.2cm}
\caption{Ground-state shell correction obtained for HF and FRDM single-particle energies using macroscopic-microscopic approach \cite{Garcia1999}.}\label{fig:shell}
\end{center}
\end{figure} 

\begin{equation}
 \rho_{Bethe}(E^*)=\frac{\sqrt{\pi}}{12} \frac{e^{2\sqrt{aE^*}}}{a^{1/4}E^{*5/4}},
 \label{bethe}
\end{equation}
where $a=(\pi^{2}/6)g$ and $E^*$ is the excitation energy. The parameter $g$ is density of single-particle states expected to be proportional to mass number $A$, leading to the result that $a$ is also proportional to $A$. It is clearly seen from Eq. (\ref{bethe}) that the logarithmic level density becomes nearly mass and excitation energy independent if it is scaled by a factor of $\sqrt{AE^*}$. In Fig. \ref{fig:hf}, we plot $\frac{ln \rho}{\sqrt{AE^*}}$ as a function of $Z$ varying around $Z_0$ for the fixed masses $40 \le A \le 180$ at $E^*=10$ and $30$ MeV. For $A=40,60,80,100,120,140,160,180$ the values of $Z_{0}$ obtained from semi-empirical mass formula are $19,27,36,44,51,59,66,74$, respectively. We also show the similar results obtained from Bethe's formula with $(Z-Z_0)$ dependent level density parameter $a$, as proposed in Ref. \cite{AlQuraishi2003}, referring it as AGMR form hereafter,
\begin{eqnarray}
a=\frac{\alpha A}{e^{{\gamma} (Z-Z_0)^2}},
\label{ldzo}
\end{eqnarray}
with $\alpha$ = 0.1068 MeV$^{-1}$and $\gamma$ = 0.0389. For a given $A$, AGMR form of $a$ displays inverted parabolic trend in the NLDs with the maximum at $Z=Z{_0}$. For $Z$ in the vicinity of $Z{_0}$, the variation of NLDs for HFB model is qualitatively similar to the ones obtained from AGMR form for $A=80$ and 100. However, for the HF model, there seems no clear maximum at $Z=Z_0$ for all the cases considered. As we go away from the $\beta$-stability line or $Z_0$, the parameter $a$ of Eq. (\ref{ldzo}) yields sharper decrease in the value of NLDs compared to those obtained for HFB model. It may be pointed out that the values of $\alpha$ and $\gamma$ in Eq. (\ref{ldzo}) were obtained by fitting the available experimental data on the NLDs for $|Z-Z{_0}| \le 1$. The fitted value of $\gamma$ is responsible for the strong suppression of the NLDs when extrapolated to $|Z-Z{_0}|>1$.\par
\begin{figure}[!htb]
\vspace{-0.3cm}
\begin{center}
\includegraphics[width=4.5in, height=3.5in]{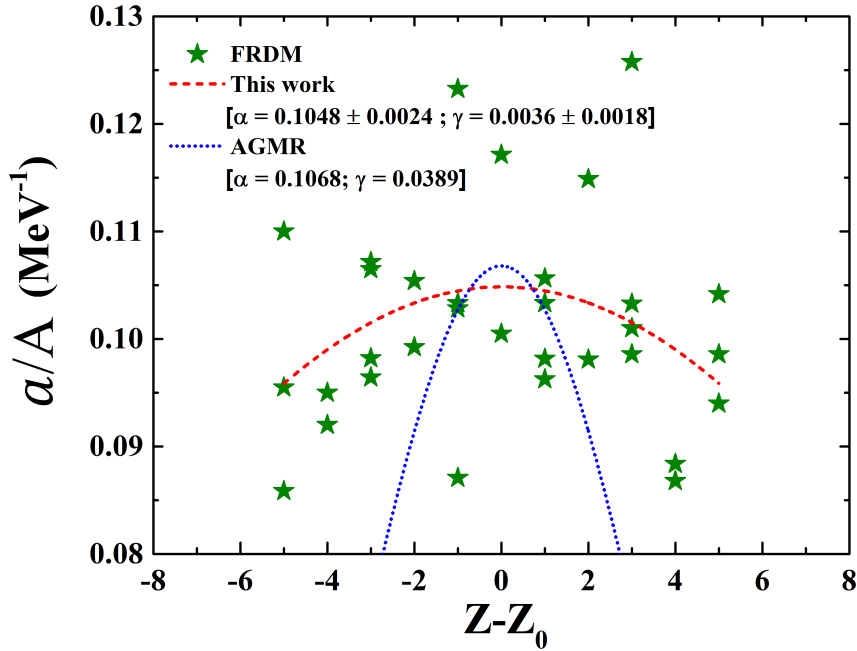}
\vspace{-0.3cm}
\caption{The level density parameter $a$ scaled by $A$ as a function of $Z-Z_0$ obtained from the NLDs using FRDM single-particle energies for a range of masses $A=40-140$. The values of scaled level density parameter are fitted to AGMR form, Eq. (\ref{ldzo}) which yield $\alpha=0.1048\pm0.0024$ MeV$^{-1}$ and $\gamma=0.0036\pm0.0018$ labelled by `This work'.}\label{fig:fit}
\end{center}
\end{figure} 

We further study the sensitivity of various microscopic structural effects on the NLDs. We compute the NLDs, by ignoring collective enhancement and pairing effects, within the statistical model \cite{Ignatyuk1993} at $E^*=30$ MeV using two different sets of single-particle energies obtained for HF and FRDM. These results are shown in Fig. \ref{fig:frdm} by triangles, and labelled as `with shell correction'. For $E^*=30$ MeV, the temperature $(T)$ is in the range of $1-2$ MeV at which the shell and pairing effects become significantly small. However, the shell effects for the ground state ($T=0$) are important as they will influence the value of $E^*$ at a finite temperature. The calculations are further performed by ignoring these shell effects at $T=0$ which are estimated using macroscopic-microscopic approach \cite{Garcia1999}. These NLDs labelled as `without shell correction' and shown by diamonds in Fig. \ref{fig:frdm}. The trends of NLDs from HF and FRDM single-particle energies with shell correction are in a qualitative agreement with the predictions of AGMR form, supporting a peak at $Z=Z_0$, for most of the cases. However, the peak is shifted little right beyond $Z_0$ for $A=140$. There is no such peak found for $A=160$ and 180. The peak structure in HF-NLDs are now somewhat visible that are absent in Fig. \ref{fig:hf} obtained from HF database which includes the collective enhancements and pairing effects. Once the shell corrections are removed, the $(Z-Z_0)$ dependence of the NLDs \cite{AlQuraishi2003} completely disappears and both the HF and FRDM models yield very similar trends. The values of ground state shell correction energies are also plotted in Fig. \ref{fig:shell}. The trends of these shell correction energies seem to play important role in $(Z-Z_0)$ dependence of the NLDs. Interestingly, the peak structure of NLDs (triangle symbols) in Fig. \ref{fig:frdm} resembles with those of shell correction in Fig. \ref{fig:shell}. These investigations imply that the $(Z-Z_0)$ dependence of NLDs is quite sensitive to the shell, pairing and collective enhancement effects. The collective enhancement and pairing effects tend to weaken the peak structure at $Z \sim Z_0$ whereas the shell effects favor it.  \par

We obtain level density parameters corresponding to the FRDM-NLDs using $a=S^2/4E^*$ for a wide range of masses at $E^*=30$ MeV. In Fig. \ref{fig:fit}, we plot the variation of level density parameter $a$ scaled by mass number $A$ as a function of $(Z-Z_0)$ for $A=40-140$. These values are fitted to the AGMR form to obtain the values of $\alpha$ and $\gamma$. We find $\alpha=0.1048\pm0.0024$ MeV$^{-1}$ and $\gamma=0.0036\pm0.0018$ from the current fit. A sizeable error in estimating $\gamma$ is due to the large spread in the values of level density parameter for a given $(Z-Z_{0})$. Our central value for $\alpha$ is very close to AGMR value, whereas, the central value for $\gamma$ is about an order of magnitude smaller than that of AGMR value \cite{AlQuraishi2003}, suggesting a weaker dependence on $(Z-Z_0)$. The reduction in $\gamma$ for heavier masses is also suggested by Grimes $et$ $al.$ \cite{grimes2008}.

\section{Conclusion}

We study the sensitivity of microscopic effects such as collective enhancement, pairing and shell corrections on the variation of total nuclear level densities in and around the $\beta$-stable nucleus $Z_0$, for a fixed mass number. For this purpose, the level densities for mass range $A=40-180$ which include these effects are taken from popularly used databases based on two different microscopic mass-models. We have also computed the NLDs using statistical model with single-particle energies from microscopic (HF) and macroscopic-microscopic (FRDM) mass-models by ignoring various microscopic effects as mentioned. We find that NLDs which include the microscopic effects show a very mild $(Z-Z_0)$ dependence compared to the inverted parabolic behavior as obtained from AGMR form of the level density parameter defined in Eq. (\ref{ldzo}). Once the effects of collective enhancement and pairing are ignored, the $(Z-Z_0)$ dependence of the NLDs becomes somewhat visible. The strongest dependence is found for $A=40-140$ using FRDM single-particle energies and yet it is weaker than that of AGMR. We also fit the level density parameter $a$, corresponding to the FRDM-NLDs, to the AGMR form. It is found that the parameter $\gamma$, which governs the exponential suppression leading to the inverted parabolic trend of level density parameter with a peak at $Z_0$, is smaller by an order of magnitude. We find that the $(Z-Z_0)$ dependence of NLDs is predominantly governed by the shell effects. The exclusion of shell effects washes out this dependence. The nature of shell effects is also sensitive to the spin-orbit interactions \cite{Kaur2015,Dwivedi2019} which changes drastically as one goes away from the line of $\beta$-stability. It may also be important to analyze the influence of spin-orbit interaction on the $(Z-Z_0)$ dependence of the level density parameter. 
\section*{Acknowledgements}
\noindent
TG acknowledges Council of Scientific and Industrial Research (CSIR), Government of India for fellowship Grant No. 09/489(0113)/2019-EMR-I. BM would like to acknowledge the financial support in the form of institute post-doctorate fellowship from IIT Ropar.

\section*{REFERENCES}

\bibliography{iopart-num}
\end{document}